\documentclass[sigconf]{acmart}
\usepackage{subcaption}
\usepackage{xcolor}
\usepackage{multirow} 
\usepackage{enumitem}
\usepackage[table]{xcolor}
\definecolor{tableShade}{gray}{0.9}
\setcopyright{none}
\renewcommand\footnotetextcopyrightpermission[1]{}

\newif\ifanonymise
\anonymisetrue       
\anonymisefalse    

\AtBeginDocument{%
  \providecommand\BibTeX{{%
    \normalfont B\kern-0.5em{\scshape i\kern-0.25em b}\kern-0.8em\TeX}}}

\begin{document}

\title{PRIMMDebug: Teaching Secondary School Students a Reflective Approach to Debugging}

\author{Laurie Gale}
\affiliation{%
  \institution{Raspberry Pi Computing Education Research Centre University of Cambridge}
  \city{Cambridge}
  \country{UK}}
\email{lpg28@cst.cam.ac.uk}

\author{Sue Sentance}
\affiliation{%
  \institution{Raspberry Pi Computing Education Research Centre University of Cambridge}
  \city{Cambridge}
  \country{UK}}
\email{ss2600@cst.cam.ac.uk}

\begin{abstract}
Debugging is a challenging and infuriating experience for many secondary school students learning their first text-based programming language. One frequent problem is the lack of reflection in students' debugging strategies, which makes error resolution unlikely and teacher reliance common. Tools that encourage more reflective and teacher-independent debugging may foster more success with fixing errors, but are lacking. This paper presents PRIMMDebug, an approach for teaching the debugging process to secondary school students. PRIMMDebug consists of an online tool that takes students through the steps of a pedagogical process based on PRIMM, a framework for teaching programming. The tool encourages written articulation throughout the debugging process and limits students' ability to run and edit code at certain stages. A classroom study with PRIMMDebug found a general reluctance among students to engage with the reflection it promotes, despite teachers appreciating this emphasis on reflection. We end by suggesting three considerations for future pedagogical debugging research and tooling: balance structure and flexibility, teach shorter debugging heuristics, and use tooling early on in students' programming journey.
\end{abstract}



\maketitle

\pagestyle{plain} 

\section{Introduction}
Debugging is a struggle for many school students learning a text-based programming language. One common problem is the lack of reflection with which many students go about finding and fixing errors. Instead of first considering the issue(s) with a program, students typically make frequent changes soon after encountering a failure \citep{MAADSMethod, ProblemSolvingDebuggingK68, ElementaryGameBasedDebugging, TeacherDiagnosticInterventionSkills}. Not only are students unlikely to consistently resolve errors with such an approach, but they may also add them to their programs \citep{ConditionsLearningNovices, ComparisonDebuggingBehaviourNoviceExpert, MAADSMethod, ThinkAloudNoviceDebugging, DebuggingNoviceSkilledProgrammers} or revert corrective changes \citep{ProblemSolvingDebuggingK68}. Struggling to fix errors can elicit feelings of frustration, surprise, and even anxiety that further impair a student's ability to debug \citep{EDAProgrammingEmotions, ProgrammingAssignmentsEmotionalToll, AnxietyLearningToProgram, EngineeringStudentsProgrammingEmotions}. If several students experience such difficulties in a classroom, teachers quickly become overloaded with struggling students to support \citep{HandsUpProblem, LackPerserveranceAutonomyDebugging, CurrentPerspectivesDebugging, StumpTheTeacher}.

To combat this lack of reflection, explicitly teaching effective debugging strategies is vital. One commonly investigated approach has been to model a \textit{systematic debugging process} (e.g., \citep{CarverImprovingChildrensDebugging, SystematicProcessMichaeli, TeachingExplicitProgrammingStrategies, DebuggingStudentsDebuggingProcess, AnalysingIntroductoryDebuggingProcess, ExplicitlyTeachingDebuggingPrimarySchool, DebuggingContextCriticalThinkingEmotion}). Following a systematic process is a global debugging strategy \citep{TeachingDebuggingFramework} that any programmer can use in any environment, and should involve explicit articulation throughout \citep{WhyProgramsFail, DebuggingByThinking}. This reflection is also important for fostering effective, rather than overreliant, use of generative AI (genAI) tools when debugging.

Although initial studies of systematic processes have reported improved debugging ability and self-efficacy \citep{CarverImprovingChildrensDebugging, SystematicProcessMichaeli, DebuggingContextCriticalThinkingEmotion, ExplicitlyTeachingDebuggingPrimarySchool}, evidence that students adopt these approaches in the long-run is lacking \citep{DebuggingInterventionLitReview}. Purely instructional approaches rely on teacher instruction, who cannot support every student in a typical school classroom, and enacting them requires self-regulation that school students are still developing \citep{TeachingExplicitProgrammingStrategies}. Students may also find them long-winded, hard to remember, or less preferable than unsystematic strategies \citep{TeachingExplicitProgrammingStrategies, NoviceReflectionsDebugging}. Pedagogical tools that guide students through a systematic process could reduce the barriers to adopting this useful strategy. While some tooling like this exists \citep{Ladebug, SystematicDebuggingLogicalErrors, ReflectiveDebuggingSpinoza}, it does not emphasise reflection throughout a structured debugging process.

In this paper, we develop a pedagogical approach based on the idea that reflecting \textit{throughout} the debugging process is a useful debugging strategy that can be effectively taught through a scaffolded tool. We present the following contributions.

\begin{itemize}
    \item We introduce \textit{PRIMMDebug} as an approach for teaching the debugging process to secondary school students learning a text-based programming language (Section \ref{sec:primmdebug}). PRIMMDebug consists of a pedagogical process and an online tool that encourages reflection throughout this process.
    \item We report two main findings from a classroom study with the PRIMMDebug tool (Sections \ref{sec:classroom-study} and \ref{sec:results}): a common student reluctance to engage with reflective debugging, and contrasting student and teacher perspectives of the tool. 
    \item Based on our findings, we list three considerations for similar pedagogical debugging tooling (Section \ref{sec:discussion}): balance structure and flexibility, teach shorter debugging heuristics, and use tooling early on in students' programming journey.
\end{itemize}

\section{Related Work}\label{sec:related-work}
Written or verbal reflection is a high-level debugging \citep{WhyProgramsFail, DebuggingByThinking} and self-regulated learning \citep{SelfRegulatedLearningAcademicAchievement, SocialCognitiveViewSelfRegulation} strategy for monitoring and controlling one's progress during the programming process. Much programming education research has focused on fostering self-regulation in students. Previous studies have explicitly taught a programming problem-solving process \citep{ProgrammingProblemSolvingSelfAwareness}, used code replays to encourage self-reflection on programming behaviours \citep{CodeReplays}, and taught effective testing practices to reduce reliance on trial and error \citep{TestingForReflectionInAction}. These interventions have improved both the quality of students' behaviours \citep{TestingForReflectionInAction, ProgrammingProblemSolvingSelfAwareness} and their awareness of those behaviours \citep{ProgrammingProblemSolvingSelfAwareness, CodeReplays}.

Most similar efforts within debugging have revolved around explicitly modelling some or parts of the debugging process (e.g., \citep{CarverImprovingChildrensDebugging, SystematicProcessMichaeli, TeachingExplicitProgrammingStrategies, DebuggingStudentsDebuggingProcess, AnalysingIntroductoryDebuggingProcess, ExplicitlyTeachingDebuggingPrimarySchool, DebuggingContextCriticalThinkingEmotion})---that is, understanding the problem and erroneous program, identifying a discrepancy between the program's intended and actual behaviour, localising the error(s), attempting and testing a fix, and repeating as necessary \citep{TeachingDebuggingFramework, AssessingDebuggingFormalModel}. Tooling that models such approaches offers \textit{metacognitive scaffolding} by helping students to monitor and control their debugging progress in a way that a purely instructional process cannot \citep{DebuggingInterventionLitReview}. Recently, generative AI tutors have been developed to this end (e.g., \citep{DebuggingAIChatbot, CompilerIntegratedConversationAIDebugging}), some of which explicitly promote articulation \citep{GenerativeAIMeetsSocraticTutor}. We now describe three tools that focus on encouraging reflection or modelling the debugging process. For a broader review of instructional debugging tooling, see \citet{TeachingDebuggingFramework} and \citet{DebuggingInterventionLitReview}.

Ladebug \citep{Ladebug} teaches debugging through a simple, constrained programming environment. Students debug erroneous teacher-created programs with the help of an omniscient debugger that enables forward and backwards stepping. Students can only edit erroneous lines of the program once they have correctly identified them, hence enforcing some adherence to the debugging process. However, Ladebug does not seem to communicate the entire debugging process to students, particularly initial program comprehension or discrepancy identification, nor does it provide anywhere for students to articulate their thoughts. The tool now appears to be inactive, and its initial evaluation only involved 14 undergraduates.

DebIso \citep{SystematicDebuggingLogicalErrors} is another tool that enacts more of a systematic process based on \citeauthor{WhyProgramsFail}'s isolation strategy \citep{WhyProgramsFail}, which involves narrowing down the cause of the error through backwards tracing and hypothesis generation. The tool contains a logbook that students can write in \textit{after} they have completed a debugging exercise, but there is no emphasis on articulating \textit{during} it. Similar to Ladebug, the evaluation of the tool only involved six undergraduates.

Lastly, Spinoza 3.0 \citep{ReflectiveDebuggingSpinoza} is a tool for encouraging reflective debugging. When students encounter an error, there is a one-in-three chance that they must write a response to two debugging prompts before being able to edit their program. This takes place with students' own programs rather than pre-written debugging exercises. In a pilot study with 101 undergraduates, \citet{ReflectiveDebuggingSpinoza} found that students who wrote reflections tended to complete problems, particularly more difficult ones, in significantly fewer steps. While these results demonstrate the utility of encouraging reflective debugging, Spinoza only contains prompts related to error identification and resolution. Encouraging reflection in the earlier stages of the debugging process may foster more success with debugging \citep{AnalysingIntroductoryDebuggingProcess}, especially for students debugging complex errors.

Overall, encouraging students to articulate their thoughts throughout the debugging process within a scaffolded tool provides metacognitive scaffolding that purely instructional systematic debugging processes lack. Some existing tools have promoted reflection or modelled some of the debugging process \citep{Ladebug, SystematicDebuggingLogicalErrors, ReflectiveDebuggingSpinoza}, but there is a lack of tooling that encourages articulation \textit{throughout} it. Evaluation of these tools with school students is also lacking.

\section{PRIMMDebug}\label{sec:primmdebug}
PRIMMDebug is an approach for teaching the debugging process to secondary school students learning to program in a text-based programming language. The approach consists of a pedagogical \textit{process} and an online \textit{tool}, both of which encourage reflective debugging. Before we discuss the tool, we detail the theoretical foundations and underlying pedagogical process it implements.

\subsection{Design Principles}

\subsubsection{Theoretical Foundations}
PRIMMDebug was underpinned by two main ideas related to reflecting throughout the debugging process: \textit{PRIMM} and \textit{systematic and scientific debugging}.

\textit{PRIMM} is a pedagogical process for teaching text-based programming \citep{PRIMMSocioculturalPerspective}. It scaffolds the process of writing code by first getting students to interact with a foreign program, supported by research on the importance of reading before writing code \citep{ListerReadingTracingStudy, ReadingWritingTracingRelationships, FurtherEvidenceReadingWritingTracingRelationships} and Vygotsky's sociocultural theory \citep{MindInSociety}. The stages of the PRIMM process are: \textit{predict} the output of a foreign program; \textit{run} the program; \textit{investigate} the structure of the program through teacher-provided activities; \textit{modify} the program's behaviour; and \textit{make} a new program with a different application to the original.

We based the PRIMMDebug process on PRIMM for several reasons. First, several of its steps are relevant to the debugging process: predicting a program's output emphasises program comprehension, running the program exposes failures in erroneous programs, and investigating code can be used to find the cause of an error. These initial debugging stages are often important for successful error resolution \citep{AnalysingIntroductoryDebuggingProcess} but frequently skipped by beginners \citep{ConditionsLearningNovices, ThinkAloudNoviceDebugging, ElementaryPuzzleBasedDebugging, ProblemSolvingDebuggingK68}. Second, PRIMM emphasises classroom talk and technical language \citep{ProgrammingTalkLanguageMediator} that likely supports intentional and correct changes when debugging. Finally, PRIMM is widely taught in English secondary schools \citep{ProgrammingSecondaryEducationEnglandWithLastAccessed, UKITSSurvey} and integrated into professional development units on programming pedagogy \citep{TeachComputingProgrammingPedagogyPDWithLastAccessed, RPFTeachPythonTeensWithLastAccessed}. Refining a pedagogical model that teachers already use lowers its barrier to classroom adoption.

\textit{Systematic debugging} is the explicit teaching of some representation of the debugging process. This approach is useful as it models the debugging stages that students often skip or struggle with. In particular, students often have difficulties with generating suitable hypotheses about the causes of errors \citep{ComparisonDebuggingBehaviourNoviceExpert, DebuggingNoviceSkilledProgrammers, AnalysingIntroductoryDebuggingProcess, TeachingExplicitProgrammingStrategies, TeachingDebuggingFramework}. This relates to \textit{scientific debugging}, which is the process of reasoning about the cause of an error using the scientific method \citep{Zeller2009ScientificDebugging}. Hypothesis generation is a key aspect of this, which \citet{Zeller2009ScientificDebugging} stresses should be done \textit{explicitly} through writing or verbalising and involve information gathering about the program and its behaviour. Although \citet{Zeller2009ScientificDebugging} suggests `quick and dirty debugging' for `simple problems', following a systematic process is a more reliable strategy when students are unable to resolve errors instantly.

\subsubsection{Guiding Principles for Development and Design} 
The theoretical foundations above informed four guiding principles for the PRIMMDebug tool's development and design.

\textbf{Principle 1: The tool should guide students through the PRIMMDebug process.} A core aspect of the tool was to provide an environment for enacting the debugging process, similar to DebIso \citep{SystematicDebuggingLogicalErrors}. This meant communicating the stages of the PRIMMDebug process and the erroneous program in the same view.

\textbf{Principle 2: The tool should promote reflective debugging by encouraging students to articulate their thoughts throughout the PRIMMDebug process.} Building on Spinoza 3.0 \citep{ReflectiveDebuggingSpinoza}, we wanted to encourage reflection throughout the entire debugging process. This could facilitate more intentional and successful changes and is in line with PRIMM's emphasis on dialogue \citep{PRIMMSocioculturalPerspective, ProgrammingTalkLanguageMediator}.

\textbf{Principle 3: The user interface should be lightweight, simple, and use clear language.} The tool is designed for secondary students in the first stages of learning a text-based programming language. It was therefore important for the user interface to only contain information that encouraged reflection throughout the debugging process. This involved guiding students' attention between a code editor and a place for writing articulations.

\textbf{Principle 4: The tool should prioritise feedback early on in the tool's deployment.} Receiving feedback on the tool, particularly from teachers, was a priority throughout development. This included frequent feedback from teachers and computing education researchers, and the classroom study reported in Sections \ref{sec:classroom-study} and \ref{sec:results}.

\subsection{The PRIMMDebug Process}
\begin{figure}
    \centering
    \includegraphics[width=0.925\linewidth]{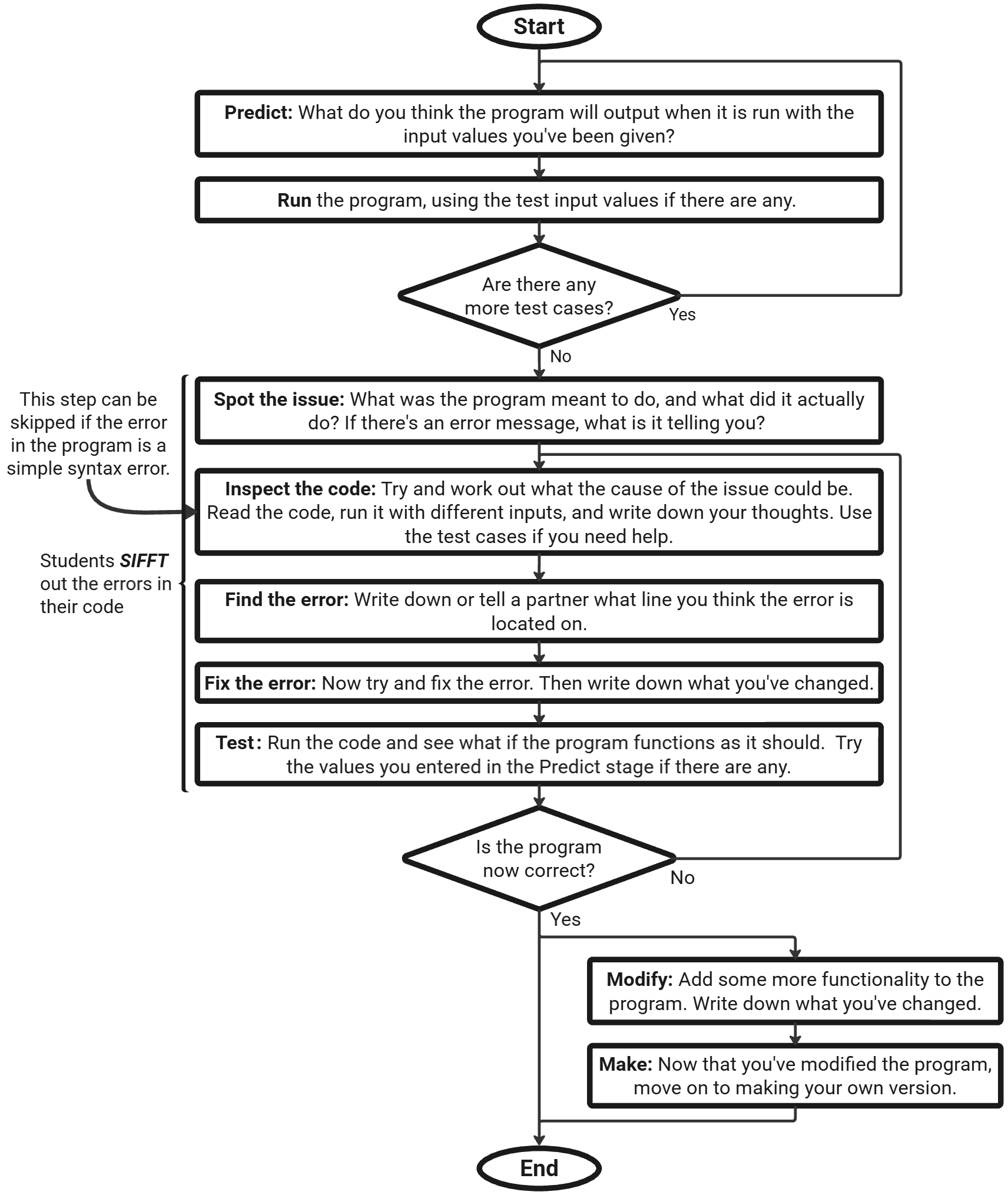}
    \caption{The PRIMMDebug process and stage prompts}
    \Description{A flowchart showing the steps of the PRIMMDebug process and the prompts for each stage.}
    \label{fig:primmdebug-process}
\end{figure}

The PRIMMDebug process is a pedagogical model for teaching the debugging process (see Figure \ref{fig:primmdebug-process}), which is enacted in the PRIMMDebug tool. It is designed for teachers to use in the classroom on PRIMMDebug \textit{challenges}, which contain the following:

    \begin{itemize}
        \item A text-based \textbf{program} with a single error that beginner programmers often make. Programs should contain one error to help students focus on learning the process \citep{DebuggingInterventionLitReview}.
        \item A \textbf{description} of the intended function of the program.
        \item Optionally, a set of \textbf{test cases} (sets of input values mapped to expected outputs). At least one of the test cases should expose the error in the program.
    \end{itemize}

The steps of the PRIMMDebug process are as follows:

\begin{itemize}
    \item \textbf{Predict:} Predict the output of the program (for a given set of input values if applicable).
    \item \textbf{Run:} Run the program and compare the program's output with the prediction, using a set of input values if applicable. If more test cases remain, begin another predict-run cycle.
    \item \textbf{Spot the issue:} Articulate the difference between the program's actual and intended output for the faulty test case(s).
    \item \textbf{Inspect the code:} Further examine the program and its behaviour to help formulate hypotheses about the nature and location of the error. In line with scientific debugging \citep{Zeller2009ScientificDebugging}, students should re-run and examine the program, re-read the description, and refine any earlier hypotheses. We discourage program editing at this stage to separate hypothesis generation from error resolution.
    
    Students should move on from this stage when they think they know the cause of the error. If students have already located the error or are dealing with simple syntax errors, this stage can be skipped.
    \item \textbf{Find the error:} Explicitly state the location of the error as specifically as possible. For example, ``an incorrect inequality in the condition of the if statement on line 3''.
    \item \textbf{Fix the error:} Attempt to resolve the error and articulate the changes made to the program.
    \item \textbf{Test:} Verify the correctness of the changes made, using the test cases for the PRIMMDebug challenge if present. If the program is still erroneous and students' changes are not trivially close to being correct, students should revert them and refine their hypotheses in the \textit{inspect the code} stage.
    \item \textbf{Modify and Make}: Once students have successfully resolved the error, they can optionally move to the less scaffolded stages of PRIMM that extend beyond the debugging process. This involves \textit{modifying} the functionality of the program, then \textit{making} a new program. In doing so, students transition to a self-created, working program.
\end{itemize}

Students are not expected to remember this entire process, nor do teachers have to explicitly communicate it to students; it is a method for completing PRIMMDebug challenges in a way that encourages reflection. However, the middle five steps, from \textit{spot the issue} to \textit{test}, are applicable to students debugging their own programs. These stages spell out the acronym \textit{SIFFT}, a systematic debugging process that teachers can encourage students to adopt when debugging their own programs. 

\subsection{The PRIMMDebug tool}
The PRIMMDebug tool is an online\footnote{\url{https://primmdebug.web.app/}}, open-source \citep{PRIMMDebugRepo} tool that takes students through the PRIMMDebug process. Like similar tooling \citep{ReflectiveDebuggingSpinoza, SystematicDebuggingLogicalErrors, Ladebug}, its key features provide metacognitive scaffolding and also emphasise reflection throughout the debugging process.

The tool consists of a landing page, a teacher information page, a dashboard listing PRIMMDebug challenges\footnote{At the time of writing, all of the PRIMMDebug challenges in the tool are in Python and were created by the first author, with several requested or refined by teachers.}, and a challenge view, where students attempt PRIMMDebug challenges.

\begin{figure}[h]
  \centering
  \includegraphics[width=0.375\textwidth]{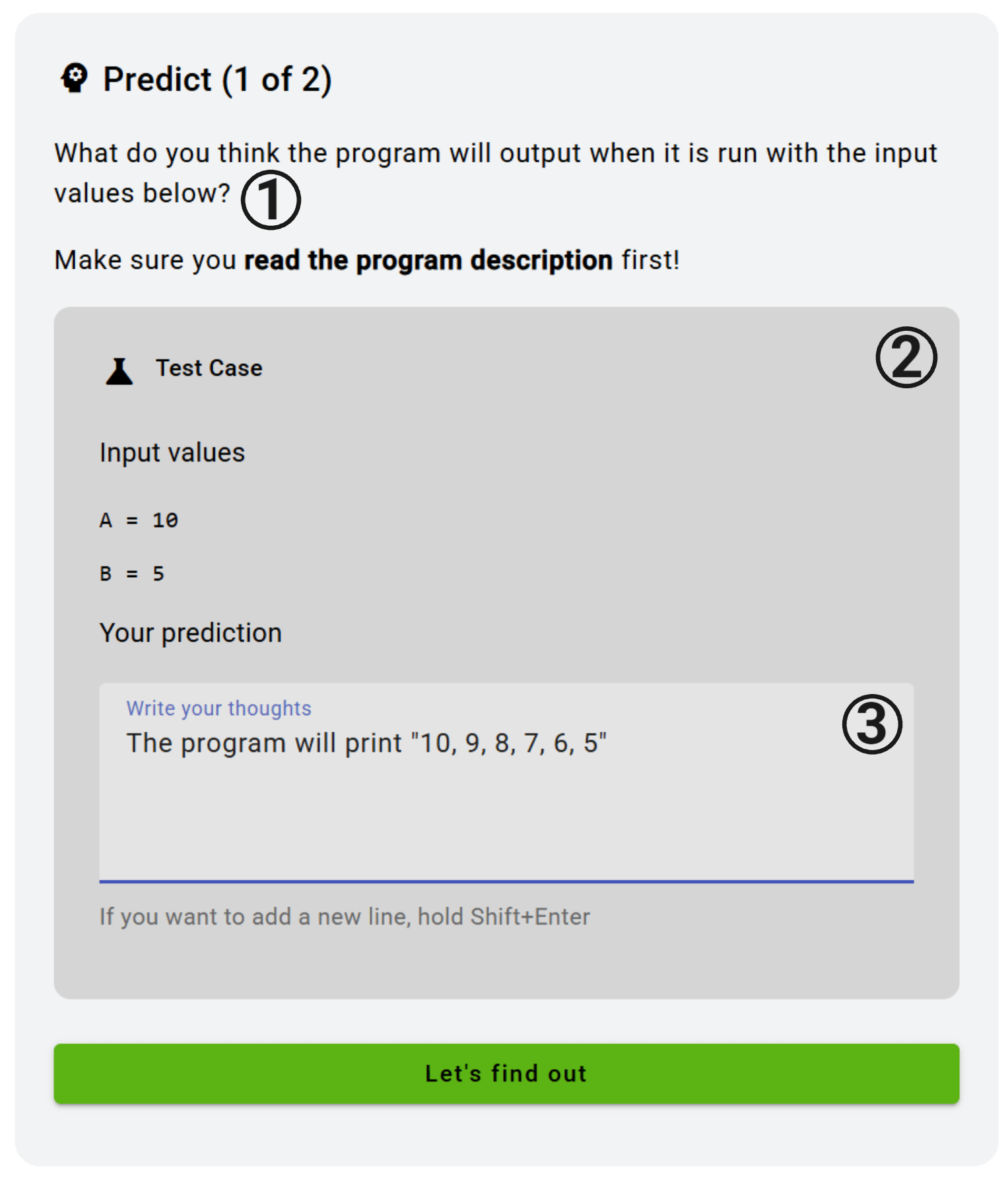}
  \caption{Articulation pane for the \textit{predict} stage, containing: \raisebox{.5pt}{\textcircled{\raisebox{-.7pt} {1}}} The PRIMMDebug stage prompt; \raisebox{.5pt}{\textcircled{\raisebox{-.9pt} {2}}} The student response pane; \raisebox{.5pt}{\textcircled{\raisebox{-.7pt} {3}}} Test case information. At this stage, students cannot run or edit the program and must write an articulation.}
  \label{fig:articulation-pane-predict}
  \Description{A small pane in the PRIMMDebug tool for the inspect the code stage, containing a stage prompt, a test case, and a text field for students to write their responses.}
\end{figure}

\begin{figure}[h]
  \centering
  \includegraphics[width=0.375\textwidth]{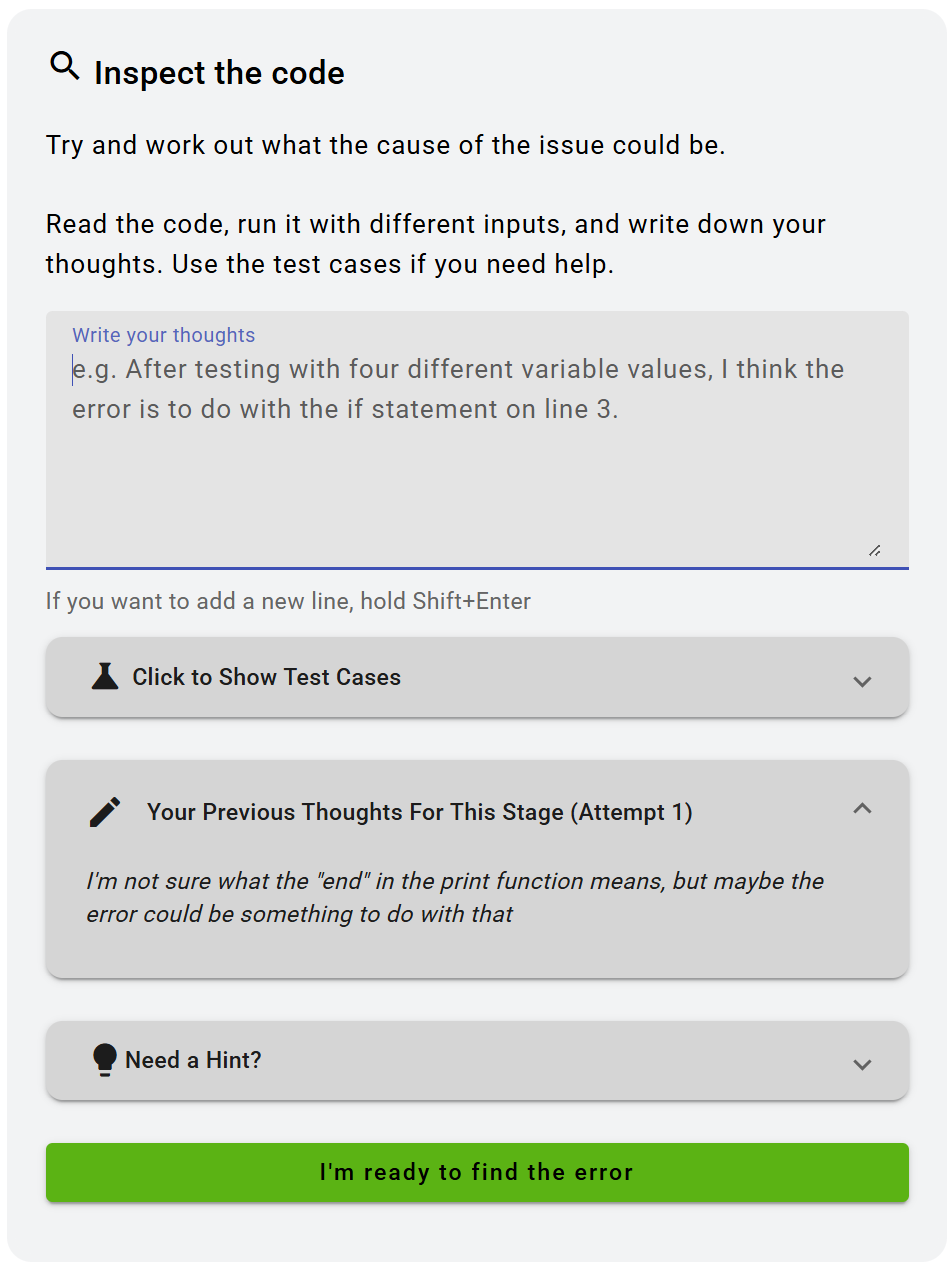}
  \caption{Articulation pane for the \textit{inspect the code} stage. At this stage, can run but not edit the program, and may write an articulation if they wish.}
  \label{fig:articulation-pane-inspect-the-code}
  \Description{A small pane in the PRIMMDebug tool for the inspect the code stage, containing a stage prompt, a dropdown for the previous stages, a dropdown for students' previous responses, a dropdown for hints, and a text field for students to write their responses.}
\end{figure}

\begin{figure}[h]
  \centering
  \includegraphics[width=0.375\textwidth]{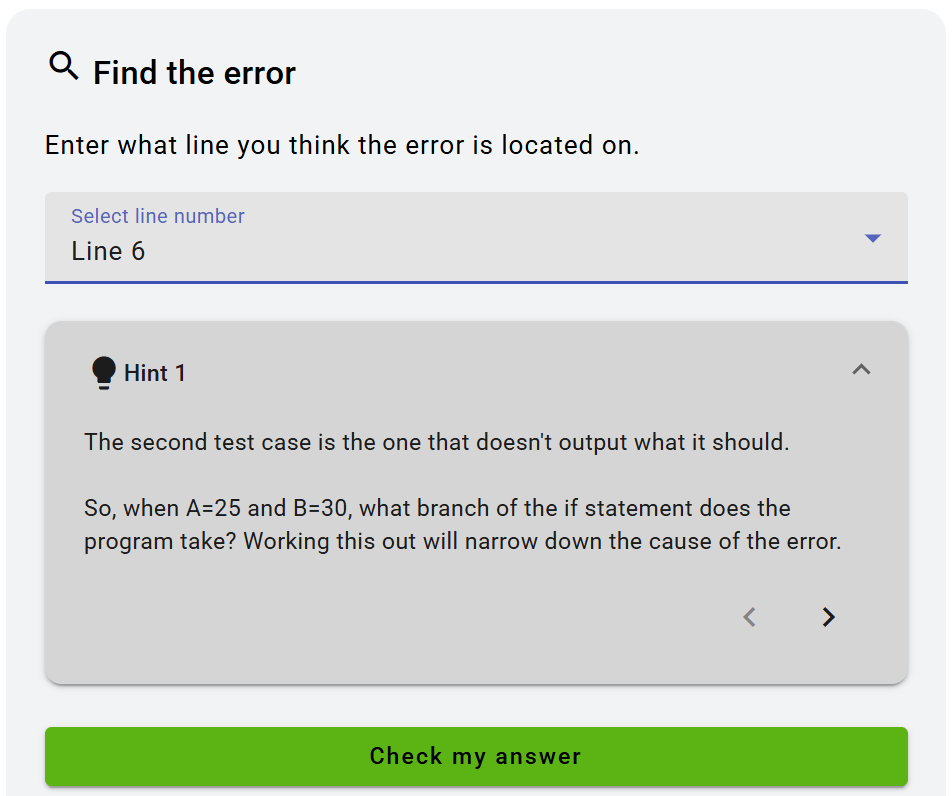}
  \caption{Articulation pane for the \textit{find the error} stage. At this stage, students must correctly select the erroneous line to move on (if the error in the PRIMMDebug challenge can be attributed to a single line).}
  \label{fig:articulation-pane-find-the-error}
  \Description{A small pane in the PRIMMDebug tool for the inspect the code stage, containing a stage prompt, a dropdown for hints, and a select field for selecting the erroneous line number.}
\end{figure}

Guided by principle 3, the UI of the challenge view is designed to be as simple as possible while conveying necessary information for encouraging reflective debugging. A simple code editor is displayed on the left of the page, the articulation pane on the right (displayed in Figures \ref{fig:articulation-pane-predict}--\ref{fig:articulation-pane-find-the-error}), and the program description at the top. The articulation pane displays information relevant to the PRIMMDebug process, including the stage prompt and test cases, and inputs students' responses.

The contents of the articulation pane change as students progress through the PRIMMDebug process. For example, test case information appears for all PRIMMDebug stages apart from \textit{fix the error}, and hints appear for \textit{inspect the code}, \textit{find the error}, and \textit{fix the error} after unsuccessful challenge attempts (see Figures \ref{fig:articulation-pane-inspect-the-code} and \ref{fig:articulation-pane-find-the-error}). Students' previous articulations are also displayed in the \textit{inspect the code} stage (see Figure \ref{fig:articulation-pane-inspect-the-code}) to aid with hypothesis refinement \citep{Zeller2009ScientificDebugging}. Each of these elements is displayed as a dropdown to avoid a cluttered display.

\subsubsection{Key Features}
In keeping with principles 1 and 2, the key features of the tool are designed to encourage reflective rather than haphazard debugging. This is primarily implemented through \textit{forced articulation}, \textit{forced localisation}, and \textit{restricted code interactivity}.

\textit{Forced articulation} involves encouraging students to write out a response to certain PRIMMDebug stages before progressing to the next one. Similar to Spinoza 3.0 \citep{ReflectiveDebuggingSpinoza}, students cannot move onto the \textit{run} stage until they have written a prediction of the program's output, or \textit{test} the program until they have described their attempted fix in the \textit{fix the error} stage. At the time of writing, students' articulations must be written and contain at least one letter or number. Allowing verbal articulation in line with PRIMM \citep{PRIMMSocioculturalPerspective, ProgrammingTalkLanguageMediator} is an important area of future work.

A feature to support successful error identification and resolution is \textit{forced localisation}. For programs where the error is located on and can be fixed by editing a single line, students must correctly identify the erroneous line before being able to fix it (see Figure \ref{fig:articulation-pane-find-the-error}). This feature is also included in Ladebug \citep{Ladebug}, which is combined with forced articulation in this tool. Correctly identifying the erroneous line does not guarantee students know the exact location of the error, but generally indicates they have narrowed down the cause.

If students do not select the correct line, they can re-attempt the \textit{find the error} or \textit{inspect the code} stage to refine their hypotheses. The next iteration of either stage will contain a hint (see Figures \ref{fig:articulation-pane-inspect-the-code} and \ref{fig:articulation-pane-find-the-error}). For challenges where the error involves several or missing lines of code, no correction checks are performed, and students instead write out their thoughts as described above. 

\textit{Restricted code interactivity} is the prevention of running or editing the program at certain PRIMMDebug stages. Students can only run the program in the \textit{run}, \textit{inspect the code}, and \textit{test} stages, and can only edit the program in the \textit{fix the error} stage. Students can run \textit{and} edit programs in the \textit{modify} and \textit{make} stages as they extend beyond the debugging process.

The acts of making and testing changes are separated to enforce this feature and limit tinkering-style debugging \citep{ConditionsLearningNovices, MAADSMethod, ProblemSolvingDebuggingK68, ElementaryGameBasedDebugging, TeacherDiagnosticInterventionSkills}. At the time of writing, students self-report the success of their fixes; if unsuccessful, they can retry the \textit{inspect the code} or \textit{fix the error} stages, and the program is reset to its original state. 

The combination of these features encourages students to understand the program they are debugging and its constituent defect before making changes to it, while limiting their ability to engage in trial-and-error-style debugging. Compared to similar tools \citep{ReflectiveDebuggingSpinoza, SystematicDebuggingLogicalErrors, Ladebug}, these features promote written articulation throughout the debugging process.

\section{Classroom Study}\label{sec:classroom-study}
We performed a mixed-methods classroom study of PRIMMDebug to gather early student and teacher perspectives and explore the extent to which students' interactions with the PRIMMDebug tool indicated adherence to the debugging process. This took place with several secondary school classes a few months into the development of PRIMMDebug and had the following research questions:

\begin{enumerate}[label=\textbf{RQ\arabic*:}]
    \item To what extent do students' interactions with the PRIMMDebug tool indicate reflection and adherence to the debugging process?
    \item What are students' perspectives on the utility of PRIMMDebug as an approach for learning about the debugging process?
    \item What are teachers' perspectives on the utility of PRIMMDebug as an approach for teaching the debugging process?
\end{enumerate}

Investigating these aims provided early insights and perspectives on reflective debugging tooling in a school classroom context, which similar tools have not been evaluated with \citep{ReflectiveDebuggingSpinoza, SystematicDebuggingLogicalErrors, Ladebug}, and informed future developments to the PRIMMDebug tool. An empirical evaluation of PRIMMDebug's effect on debugging behaviour was premature for a first study; we prioritised gathering early data to improve the tool before conducting such a study.

\subsection{Study Procedure and Data Collection}
\begin{figure*}
    \centering
    \includegraphics[width=0.825\linewidth]{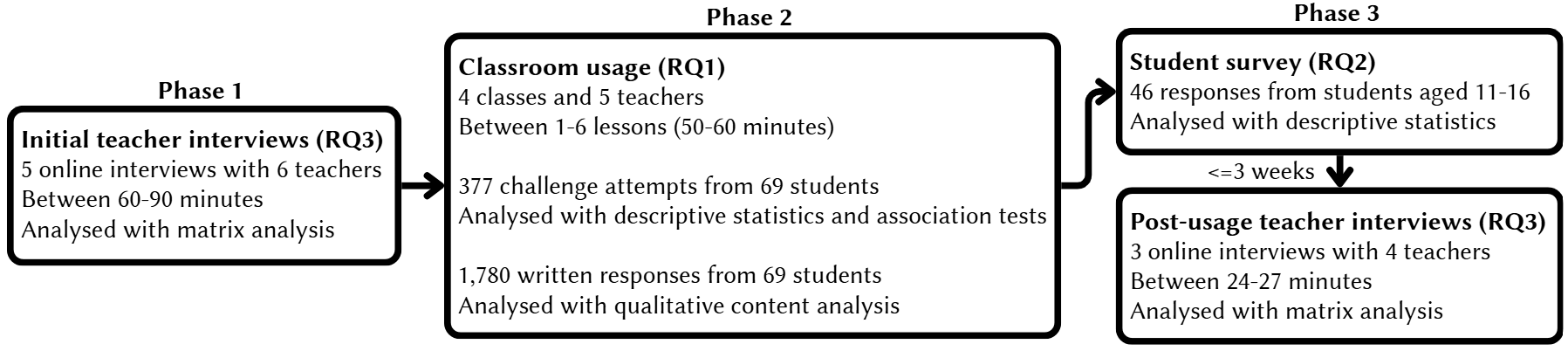}
    \caption{An overview of the PRIMMDebug classroom study} 
    \label{fig:study-overview}
    \Description{A diagram visualising the design of the PRIMMDebug classroom study. It contains three boxes of summative text, corresponding to the phases of the study described in Section 4.}
\end{figure*}

The study involved four secondary school computing teachers teaching with PRIMMDebug to their students over several lessons (see Figure \ref{fig:study-overview}). The design was refined through a small-scale pilot study with a local secondary school class in October--November 2024 and approved by our department's ethics committee (review ID \#2353). All of the study materials, research data, and analysis code are available in the study repository \citep{Study3Repository}.

\subsubsection{Initial Teacher Interviews}
Secondary computing teachers who taught Python were purposively sampled through a call in our research centre's newsletter. Six teachers agreed to participate, but only four completed all three phases of the study\footnote{One teacher only completed the first phase and another completed phases 1--2.}. Two of these teachers (Ashley and Blake) participated together with a class that they co-taught.

The first author then met with each teacher to introduce them to PRIMMDebug, let them use the tool, and gather initial qualitative perspectives on it. All teachers requested minor changes to the tool, which were acted on before the next phase of the study. Notably, most teachers requested challenges that contained syntax errors, which were not previously present in the tool. This feedback was interesting, as the entire SIFFT process would not usually be required for simple syntax errors. However, many students struggle to locate and resolve syntax errors \citep{UnderstandingSyntaxBarrier, JadudCompilationBehaviour, InvestigationSchoolStudentsDebuggingBehaviourPython}, in which case a systematic process may help to structure their efforts. More challenges with simple syntax errors were therefore added to facilitate classroom usage for students of all abilities.

Interviews lasted between 60--90 minutes and were recorded and automatically transcribed using Microsoft Teams or Google Meet. The first author checked the transcripts to correct transcription errors, remove personal information, and write initial memos. These anonymised transcripts were then sent to the teachers within four weeks of the interviews, who were invited to redact or correct anything they had said.

\subsubsection{Classroom Use}
Teachers then taught with PRIMMDebug in several of their computing lessons. The exact number varied, as we gave teachers freedom over how and how long they taught with PRIMMDebug. We only asked that students attempt at least one PRIMMDebug challenge in each lesson that teachers taught with it. We also provided teachers with additional resources to support their classroom practice, including a poster for the SIFFT process.

To investigate RQ1, we logged the following data from students' usage of the PRIMMDebug tool:
\begin{itemize}
    \item On completing a PRIMMDebug stage: The written articulation (if applicable) and a timestamp of when the stage was completed. For the \textit{find the error} stage, the correctness of students' responses was also logged.
    \item On running a program: A program snapshot, a timestamp of when it was run, any I/O and error messages from the run, and whether the program successfully executed.
\end{itemize}

We only collected data from students who consented to participate by using a separate `research' version of the tool that collected log data and required a login only provided to consenting students.

\subsubsection{Post-Usage Perspectives}
After classroom usage, teachers were interviewed and students were surveyed. The student survey was developed by the authors to gather quantitative feedback. It asked about the usability of the tool and the perceived helpfulness of its key features and the SIFFT process for learning about debugging. We report on the latter two items in Section \ref{sec:students-perspectives}, which were answered on a Likert scale of 1--4. The survey was hosted on Qualtrics and was accessible in the research version of the tool.

In the post-interviews, teachers shared how they taught with PRIMMDebug, their observations of their students' experiences with it, and their qualitative perspectives on the effectiveness of the approach. The same procedure as the initial interviews was used to conduct, record, and transcribe the interviews.

\subsection{Participants}
Four teachers completed all three phases of the study (see Table \ref{tab:teacher-demographic-breakdown} for demographic information). Teachers received two \pounds25 bookshop vouchers for participating---one after the initial interview and another after the post-interview. Out of the schools that teachers taught at, one was a high-performing, selective, boys-only grammar school, another was a private international school, and the other was a mixed-gender state-funded school. Of the classes that teachers taught PRIMMDebug with, one was grade 6 (ages 11--12), where computing was a mandatory subject and students were learning Python for the first time, and the other two were grade 9 (ages 15--16), where students had opted to study computer science and had approximately 1--2 years of Python experience. All students also had several years of block-based programming experience as part of their primary computing education \citep{UKPrimaryComputingCurriculumWithLastAccessed}.

\begin{table}[h]
 \caption{Teachers' demographic information}
 \label{tab:teacher-demographic-breakdown}
    \begin{tabular}{p{2.75cm} p{5.25cm}}
        \toprule
         \textbf{Characteristic} & \textbf{Demographic breakdown} \\
         \midrule
         Gender & Male: 2 \hspace{0.25cm} Female: 2 \\
         Age & 18--29: 2 \hspace{0.25cm} 50--59: 1 \hspace{0.25cm} Over 60 years: 1 \\
         Years of computing teaching experience & 0--5: 2 \hspace{0.25cm} 16--20: 1 \hspace{0.25cm} Over 20 years: 1 \\
         \bottomrule
    \end{tabular}
\end{table}

69 students used the research version of the tool, but only 46 students completed the survey, as one teacher taught with PRIMMDebug but did not complete the final phase of the study. Students self-reported their year group and gender (from male, female, other, or prefer not to say) in the post-survey, of which 37 reported male, 7 female, and 2 preferred not to answer.

\subsection{Analysis}

\subsubsection{Matrix Analysis of Teachers' Interviews}
Teachers' interviews were analysed with matrix analysis \citep{MilesHubermanQualitativeDataAnalysis, UsingDataMatrices}, a qualitative analysis method that uses coding cycles to cross-tabulate two aspects of a dataset. We developed a \textit{conceptually clustered matrix}, which tabulates individual participants against high-level categories \citep{MilesHubermanQualitativeDataAnalysis}, due to the number of teachers and data sources being analysed.

The analysis was conducted by the first author in ATLAS.ti using several interleaved stages. First, the interview transcripts were read through, during which memos were written and descriptive open codes related to RQ3 were generated in a process of \textit{first cycle coding} \citep{MilesHubermanQualitativeDataAnalysis}. By re-reading the transcripts and discussions between the authors, open codes were merged, deleted, and refined into four high-level categories in a process of \textit{second cycle coding}. Each category contained two to four sub-categories that are not reported in the matrix but were used to facilitate the generation of it.

To verify that the codebook was clear and could be consistently applied by the first author, otherwise known as \textit{intracoder reliability} \citep{IntercoderReliabilityQualAnalysisGuidelines}, agreement checks were performed with a research colleague unfamiliar with the data. After introducing them to the study and analysis procedure, we independently coded a randomly selected post-interview transcript (Drew). We then met to identify coding discrepancies and refine the codebook in a process of \textit{negotiated agreement} \citep{UnitisationProblemSemiStructuredInterviews}. This was preferable to calculating an \textit{intercoder reliability} coefficient due to the length of the transcripts and the difficulty in agreeing on exact units of analysis \citep{UnitisationProblemSemiStructuredInterviews}. Finally, all of the interviews were coded with the final codebook and the matrix was developed to summarise each teacher's perspective.

\subsubsection{Analysis of Students' Survey Responses}
The distributions of survey items related to the perceived helpfulness of the tool's key features and the SIFFT process are plotted and triangulated with students' interactions (RQ1) and teachers' perspectives (RQ3) throughout Sections \ref{sec:results} and \ref{sec:discussion}. The internal consistency of the items related to the tool's key features, as measured by Cronbach's alpha ($\alpha$), was 0.72, suggesting reliable items \citep{CohenDescriptiveStatsChapter}.

\subsubsection{Content Analysis of Students' Written Articulations}
Students wrote a total of 1,780 articulations into the tool. We used qualitative content analysis \citep{QualContentAnalysis} to summarise 900 (51\%) of these with inductive codes, stratified by PRIMMDebug stage, which were then abstracted into levels of \textit{adherence} to the PRIMMDebug process. Adherence represents the closeness of a student's written articulation to the PRIMMDebug stage prompt they were responding to, and was split into the three levels below.

\begin{itemize}
    \item \textbf{Full adherence:} Direct answers to the prompt that students were writing a response to. For example, articulating the difference between the actual and intended behaviour of the program in the \textit{spot the issue} stage.
    \item \textbf{Partial adherence:} Codes that did not completely answer the stage students were attempting but were of wider relevance to the PRIMMDebug process. For example, identifying the cause of the error in the \textit{predict} stage.
    \item \textbf{No adherence:} Written articulations that were not relevant to the PRIMMDebug process, such as random strings of text.
\end{itemize}

The content analysis used for analysing written adherence had several stages. First, the first author manually inspected students' written articulations and inductively developed codebooks for the four PRIMMDebug stages where students could write a response (see Table \ref{tab:student-adherence-categorisation}). These codebooks were refined through three coding rounds with different research colleagues on samples of 90 responses (5\%) stratified by PRIMMDebug stage. To evaluate the codebooks' intercoder reliability (ICR), the authors then performed several joint coding rounds, each consisting of 180 articulations (10\% of total responses) stratified by PRIMMDebug stage. We used Krippendorff's alpha ($\alpha$) \citep{ReliabilityContentAnalysisBook} as the reliability coefficient, and performed coding rounds until each codebook's $\alpha$ exceeded \citeauthor{ReliabilityContentAnalysisBook}'s recommended threshold of 0.8 (see Table \ref{tab:student-adherence-categorisation}). The first author then coded 51\% of students' written articulations ($n = 900$), also stratified by PRIMMDebug stage, including 80 articulations that could not be reliably coded as one category and were discounted from analysis. Finally, we assigned each code to one of the adherence levels above, which are reported in Table \ref{tab:student-adherence-categorisation}. Adherence was treated as an ordinal variable---the more adherent a code, the more detailed and high-quality the written articulations it represents.

We then compared students' articulation adherence with the success of their challenge attempts\footnote{A binary variable for whether a student's final program passed a set of test harnesses. These were written during analysis and used instead of students' self-reported success.} using Cochran-Armitage tests for trend ($T$), and the time spent on each PRIMMDebug stage using Kendall's tau ($\tau$) tests (due to the non-normal distribution of time observed from visual inspection). $p$-values less than 0.05 are considered statistically significant.

The adherence variable naturally had some limitations. Some students may have struggled to write detailed articulations but reflected extensively, while others may have written short articulations with enough keywords to be categorised as fully adherent. However, there was no easy way to mitigate these issues without significant context and interpretation, which would have likely reduced the reliability of the codebooks.

\section{Results}\label{sec:results}
\subsection{Students' Interactions (RQ1)}\label{sec:student-interactions-primmdebug}
\begin{figure}
    \centering
    \includegraphics[width=0.975\linewidth]{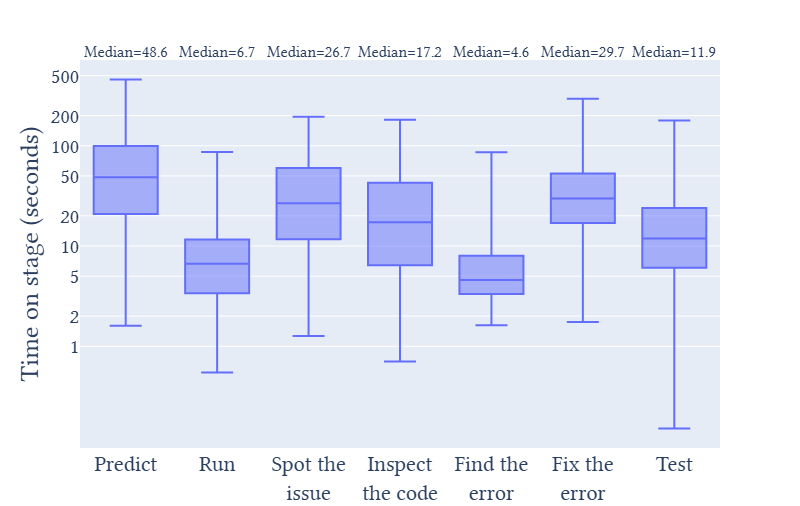}
    \caption{The median time spent on each PRIMMDebug stage}
    \label{fig:time-per-primmdebug-stage}
    \Description{A box plot visualising the log time spent on each PRIMMDebug stage.}
\end{figure}

Students attempted a total of 377 PRIMMDebug challenges. As one teacher only used PRIMMDebug in one lesson, 186 (49\%) of these took place in the first session. The three most attempted challenges (47\% of attempts) contained a syntax error, and 58\% of attempts did not successfully resolve the erroneous programs. The median time per challenge attempt was 5 minutes and 46 seconds ($skew = 2.06$), which is broken down by PRIMMDebug stage in Figure \ref{fig:time-per-primmdebug-stage}. 

Table \ref{tab:student-adherence-categorisation} reports the adherence of students' written articulations for each PRIMMDebug stage where students could write a response\footnote{The varying number of articulations for each stage is due to the structure of the PRIMMDebug process in the tool. Students usually had to go through multiple \textit{predict} stages, but did not have to enter a response for \textit{inspect the code}.}, which was statistically significantly correlated with the time per stage ($\tau = 0.27, p < 0.001$) with a small effect size\footnote{As per \citeauthor{StatisticalSignificanceEffectSizes}'s guidelines \citep{StatisticalSignificanceEffectSizes}.}. We now report students' interactions by PRIMMDebug stage.

\begin{table*}[h]
 \caption{The adherence of students' written articulations in the PRIMMDebug tool}
 \centering
 \label{tab:student-adherence-categorisation}
    \begin{tabular}{c p{2.5cm} p{2.5cm} p{2.5cm} p{2.5cm} c}
        \toprule
        & \textbf{\textit{Predict} ($n=341, \alpha = 0.824$)} & \textbf{\textit{Spot the issue} \newline ($n=156, \alpha = 0.895$)} & \textbf{\textit{Inspect the code} ($n=108, \alpha = 0.819$)} & \textbf{\textit{Fix the error} \newline ($n=215, \alpha = 0.856$)} & \textbf{Total ($n=820$)} \\
        \midrule
        Full adherence & 62\% (211) & 24\% (38) & 54\% (58) & 29\% (63) & 45\% (370) \\
        Partial adherence & 19\% (64) & 62\% (96) & 30\% (32) & 47\% (102) & 36\% (294) \\
        No adherence & 19\% (66) & 14\% (22) & 17\% (18) & 23\% (50) & 19\% (156) \\
        \bottomrule
    \end{tabular}
\end{table*}

\subsubsection{Predict and Run}\label{sec:predict-run-behaviour}
Students could not run or edit the erroneous program in the \textit{predict} stage. Instead, they were instructed to predict the program's output for a given test case. Students wrote the highest proportion of fully adherent articulations and generally spent the longest time on this stage, of which there were usually multiple. Among the fully adherent articulations, students mostly predicted the program's output (131 responses, 38\%) as instructed or summarised the program's output or behaviour in plain English (73, 21\%). Teachers appreciated how the forced articulation at this stage encouraged program comprehension (see Section \ref{sec:teachers-perspectives}). In the following \textit{run} stage, students were able to run but not edit the program. This was the least time-consuming stage, taking a median of 6.73 seconds ($skew = 10.89$).

\subsubsection{Spot the Issue}
Students could not run or edit the program at this stage and were instructed to articulate the difference between the program's actual and intended behaviour for the failing test case. This stage contained the least fully adherent written articulations (24\%) across the PRIMMDebug process. An equal number of articulations (33, 21\%) contained perceptions of the program's actual \textit{and} expected behaviour (fully adherent) or the program's actual \textit{or} intended behaviour (partially adherent). Other partially adherent written articulations mentioned the error message thrown (28, 18\%) or related to a different PRIMMDebug stage (27, 17\%).

\subsubsection{Inspect the Code}\label{sec:inspect-code-behaviour}
Students were encouraged to narrow down the cause of the error by running the code with different inputs, re-reading the program code and description, and optionally writing any thoughts. In practice, students did not write anything in 45\% of attempts, did not run the program in 77\% of attempts, and spent a median of 19 seconds ($skew = 5.64$) on this stage. For students who did write an articulation, there was a statistically significant association between the adherence of their articulations and success in the following \textit{find the error} stage ($T = 54, \ n = 110 \ p = 0.008$)---students who wrote more adherent responses were generally more successful. Common written articulations hypothesised the cause of the defect (34, 31\%), suggested potential fixes (19, 18\%), or simply stated a line number (21, 19\%).

\subsubsection{Find the Error and Fix the Error}\label{sec:student-engagement-find-the-error}
Students had to correctly identify the erroneous line of the program (if possible) before moving on from the \textit{find the error} stage. This was successfully done in 255 of 308 first attempts (83\%), although 42\% of challenge attempts did not successfully resolve the error in the \textit{fix the error} stage.

Students also had to write the changes they made in the \textit{fix the error} stage. The adherence of these articulations was significantly related to the correctness of students' changes ($T = 78, \ n = 169, \ p = 0.022$). Students who wrote more adherent articulations had a higher proportion of correct changes. Among these responses, 63 (29\%) were detailed and reproducible descriptions of changes (fully adherent), while 94 (44\%) were vague, unreproducible summaries (partially adherent).

\subsubsection{Test}
At the time of the evaluation, students were not provided with test harnesses. They were instead encouraged to test the programs themselves and self-report the correctness of their changes. However, students only spent a median time of 12 seconds testing and generally did not run the program at all (26\% of attempts) or only ran it once (57\% of attempts), despite the majority of challenges containing multiple test cases. 

\subsection{Students' Perspectives (RQ2)}\label{sec:students-perspectives}
\begin{figure}
    \centering
    \includegraphics[width=0.975\linewidth]{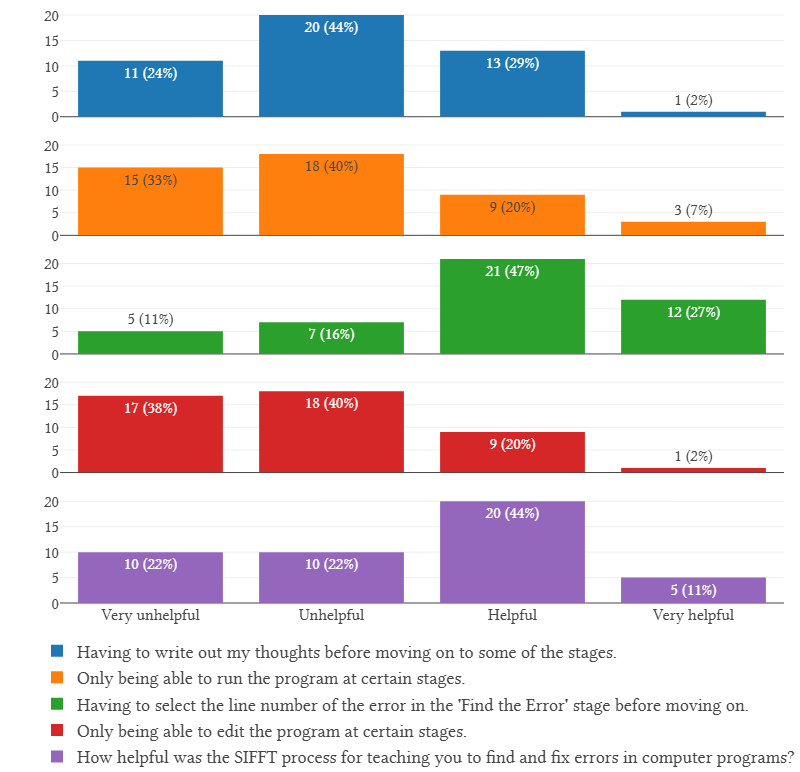}
    \caption{Students' perspectives on the helpfulness of the PRIMMDebug tool's key features and SIFFT process}
    \label{fig:survey-responses}
    \Description{A set of bar charts visualising students' perspectives towards the key features of the PRIMMDebug tool and the helpfulness of the SIFFT process}
\end{figure}

Most students reported the tool's key features as unhelpful for learning about debugging (see Figure \ref{fig:survey-responses}). The only exception was the positive perspectives towards the forced localisation feature, which aligns with students' high levels of success on the \textit{find the error} stage. Students' perspectives of the SIFFT process were mixed (see Figure \ref{fig:survey-responses}). The number of students who found it helpful was equal to those who found it unhelpful or very unhelpful (20, 44\%). Despite the lack of success, 21 students (47\%) reported the challenges to be a `good level of difficulty' in the survey, and only 6 (13\%) reported them as too challenging.

\subsection{Teachers' Perspectives (RQ3)}\label{sec:teachers-perspectives}
Table \ref{tab:primmdebug-teacher-matrix} presents teachers' perspectives using the four categories developed from matrix analysis, which we now summarise. First, \textbf{teachers generally appreciated the reflective approach that PRIMMDebug incorporated}. All teachers mentioned that the SIFFT process was a useful model for their own teaching practice and their students (\textit{``having a deeper understanding of their code and reflecting on what they've changed, I think this is a really cool framework for them to follow''}---Drew). All the teachers also found the key features of the tool useful for encouraging program comprehension and intentional fixes.

However, the \textbf{emphasis on reflection was counter to many students' debugging strategies}. This inability to \textit{``straight away start chopping out lines''} (Blake) translated to common frustration, aligning with students' perspectives. Contrarily, Drew reported some students appreciating the ability to \textit{``plan their attack''}, and some teachers observed \textbf{weaker students engaging with reflective debugging more}. These students appeared to articulate meaningful responses that helped them to debug, while \textbf{more confident or able students were reportedly more frustrated with the tool}. This led to suggestions for more flexibility in the tool, such as making PRIMMDebug stages skippable. Some teachers also \textbf{questioned the possibility of students adopting the SIFFT process in their own debugging practice}. Ashley cited the length of the process as a barrier to enacting it, calling for three simple steps.

Teachers also discussed future applications of PRIMMDebug. Chris was interested in teaching it to their younger students, in the hope that a systematic debugging approach would become habitual. Others suggestions mostly related to longer-term classroom usage, such as a `teacher view' to see their students' articulations.

{\rowcolors{1}{tableShade}{white}
\begin{table*}[h]
 \caption{Teachers' perspectives of PRIMMDebug}
 \centering
 \label{tab:primmdebug-teacher-matrix}
 \begin{tabular}{p{0.75cm} p{3.5cm} p{3.5cm} p{3.75cm} p{3.25cm}}
    \toprule
     & \textbf{Perspectives on the process} & \textbf{Perspectives on the tool} & \textbf{Student observations and perspectives} & \textbf{PRIMMDebug going forwards} \\
    \midrule
    Ashley & Having a framework for the debugging process was useful, but a shorter version would be more helpful for students. \newline \textit{``The steps all make a lot of sense ... but there's almost too many to quickly remember ... you almost want it to be like a three-step process rather than a five.''} & The forced articulation was useful for helping students to \textit{``formulate their answers more''}---\textit{``I think that's a good thing to force them to have to step back and think about it a little bit more.''} \newline The structuring of the process within the tool did not cater for different abilities; this should be more flexible. & \textit{``I think it depended on their own confidence and abilities and their general attitude towards coding.''} \newline Weaker students `bought in' to the rigid structuring of the process, while more confident students became frustrated with it. & More features to aid longer-term classroom usage, including teacher creation and setting of PRIMMDebug challenges. \newline These would help the tool to be used at more points where students are struggling with common errors. \\
    
    Blake & Liked the concreteness that the process provided for debugging and the emphasis on reflection, but their students find it hard to \textit{``embed any sort of routine''}. \newline \textit{``I think it's a good thing to make them slow down. I think we're not used to making them do that sometimes.''} & The tool was easy to use in the classroom. The key features were useful but counter to students' typical debugging strategies. \newline \textit{``I like it }[restricted code editability]\textit{. Given the choice, they would \dots start chopping out lines and putting these things in.''} & Weaker students generally engaged with and enjoyed using the tool more. More confident students were frustrated with not being able to fix the error straightaway. \newline Students saw the value in completing PRIMMDebug challenges with errors they may encounter. & Making the process more flexible within the tool, with more scaffolding for struggling students, would be useful. \newline \textit{``If we could sort of make our own }[challenges]\textit{ for the tool and put it on that, I can see us doing that with earlier years.''} \\
    
    Chris & The process was intuitive, adaptable to different levels of experience, and gave students a framework to use when stuck.  \newline \textit{``The single most useful thing is that it's causing children to reflect and not just jump in.''} & The tool was \textit{``really easy to implement and satisfying''}. \newline In favour of the restrictive features, as they encourage deeper thinking that students often do not engage in---\textit{``otherwise they'll simply rush in.''} & Some `brighter' students were frustrated by the key features of the tool, while others appreciated the \textit{``repetition and reinforcement''}. \newline Most students `came round' to reflective debugging---\textit{``I would say the majority of them ... thought that it was a really good idea and were happy that we would continue using it.''} & Wants to teach PRIMMDebug to their youngest students (grade six) in the hope it becomes \textit{``part of their normal method''}. \newline More PRIMMDebug challenges would help with this, as would being able to create their own. \\
    
    Drew & The process was useful for encouraging reflection, technical language, and giving students something to refer to when debugging. \newline \textit{``It’s been good to have more conversations about ... how we would debug things ... with a bit more structure.''} \newline The SIFFT process would need to be encouraged over time. & Appreciated the key features as a means for program comprehension and making intentional changes. Without this, some students would \textit{``immediately tinker''}. \newline \textit{``I think this was quite good to get them to really think about it. They had no option but to think before running the code.''} & Most students were initially frustrated with being unable to run the code straightaway, but this generally reduced over time. \newline Some students were very engaged---\textit{``they quite enjoyed having the time to think about their code without the pressure of running it straightaway. They could kind of plan their attack''}. & Would like to use PRIMMDebug challenges without telling students the program contains an error. \newline Interested in seeing students' responses to understand engagement and common struggles, and a \textit{``teacher view''} to create their own challenges. \\
    \bottomrule
    \end{tabular}
\end{table*}
}

\section{Discussion and Recommendations}\label{sec:discussion}
Many of our findings relate to the way that the PRIMMDebug tool encourages reflection throughout the PRIMMDebug process. We aimed to provide scalable modelling and metacognitive scaffolding that built on purely instructional systematic processes. However, we found challenges specific to our tool-process approach: students were often reluctant to engage with certain stages of the process, not able to fix the errors in the challenges, and frequently not in favour of the key features of the tool. The quality of students' written articulations also varied, with some relation between their adherence to the debugging process and success in the tool.

Based on our findings and related work, we list three considerations for the future design and investigation of pedagogical debugging tools. Researchers should investigate the effect of these design choices on students' perspectives, behaviours, and success with resolving errors.

\subsection{Balance Structure and Flexibility}
Some teachers felt the structure and length of the SIFFT process were barriers to its adoption, which aligns with other students' perspectives of systematic processes \citep{TeachingExplicitProgrammingStrategies, NoviceReflectionsDebugging}. Therefore, the systematic process that pedagogical tools convey must be carefully considered, and \textit{how} these processes are communicated through a tool must balance structure and flexibility. Systematic processes could be error-dependent (e.g., \citep{SystematicProcessMichaeli}), though this comes at the cost of memorability. Instead, keeping processes simple, error-agnostic, and their implementation within tools more flexible could improve student perspectives and increase their adoptability. 

\subsection{Teach Shorter Debugging Heuristics}
Another way to encourage reflective and systematic debugging is to teach shorter heuristics than a complete model of the debugging process. One teacher suggested a three-step process, while \citet[p. 474]{TeachingExplicitProgrammingStrategies} suggests a simple `find and understand the cause before editing' approach. Although these heuristics do not model the entire debugging process, they are still helpful global debugging strategies and are certainly more beneficial if students feel more able to use them. These can easily be integrated into beginner programming environments for students to use with their own errors \citep{ReflectiveDebuggingSpinoza}.

Alternatively, tools could not model any heuristic and simply encourage articulation when debugging. This approach is more aligned with how students debug using genAI tools, with recent `AI Socratic tutors' designed to encourage this conversational approach  (e.g., \citep{GoodRubberDuckDoesNotQuack, TeachingCS50WithAI, CompilerIntegratedConversationAIDebugging}). These tools will likely become more popular than those that teach an entire systematic process, as they are more flexible and adaptive. The effect of these types of tools on students' debugging behaviour should be evaluated.

\subsection{Introduce Tooling Early}
Another barrier to debugging systematically is the preference to use unsystematic strategies  \citep{TeachingExplicitProgrammingStrategies}. Some students in our study had had years to develop unreflective debugging behaviours; using a tool that discourages these likely contributed to their interactions and perspectives. Future work should use similar tooling with students at the very beginning of learning to program and investigate how their debugging behaviours develop. While many of the errors that students encounter in the first stages of learning to program will not warrant extensive reflection or systematicity \citep{SystematicProcessMichaeli}, tools that balance structure with flexibility are arguably useful for any error type. If such studies find a development of positive debugging behaviours, teachers will be incentivised to adopt such tools in their classrooms.

As more school students learn to program with genAI tools, the nature of debugging changes. GenAI tools can swiftly complete certain parts of the debugging process that students may struggle with. Similar pedagogical tools that teach AI-oriented debugging strategies should still encourage systematicity and reflection, but may have to emphasise certain parts of the debugging process more than others. For example, genAI tools can help develop hypotheses for the causes of errors and potential fixes, but students must be proficient at understanding the program and verifying fixes. Scaffolded genAI tools can also provide personalised and adaptive support, such as stage-specific hints based on students' progress.

\subsection{Limitations}
There are several limitations with the PRIMMDebug tool that may have affected students' interactions. The tool lacked some useful features at the time of the classroom study, most notably test harnesses and the ability to record articulations. We encourage similar tooling to incorporate these features. The specific wording of the stage prompts may also have been confusing for some students \citep{AnalysingIntroductoryDebuggingProcess}, though these were refined after teacher feedback.

The classroom study also had several limitations that affect the validity and generalisability of our findings. The student sample was male-biased and from a small number of classes that varied in age and motivation. Some were older students who had opted to study computer science, while others were studying computing as a compulsory subject. This would have made the PRIMMDebug challenges too easy or difficult for some students and reduced the need for some to be guided through the PRIMMDebug process. Additionally, the student survey we used was unvalidated and some students only used the tool for one lesson, which would have affected their usage and perspectives of the tool. Finally, several data were not collected that would have provided more context to the results, including students' programming self-efficacy and ability.

\section{Conclusion}\label{sec:conclusion}
We have presented PRIMMDebug, a pedagogical approach that encourages students to articulate their thoughts throughout the debugging process via an online tool. A classroom study with PRIMMDebug found that students were often reluctant to engage with the reflection tool that the tool encourages, and found most of its key features unhelpful for debugging. However, teachers appreciated the tool's focus on reflection as a means of encouraging intentional fixes. Our tool and findings represent an early effort to teach the debugging process to school students through a tool. To help students adopt a more systematic approach in the future, more flexible tooling and shorter debugging heuristics should be used early on in students' programming journey, and the effects of these tools on debugging behaviour should be investigated over multiple school terms. Researchers must also consider which other debugging strategies to model in similar tools as school students become more accustomed to debugging with genAI tools.

\begin{acks}
The authors would like to thank all of the teachers who took part in the classroom study and gave feedback on PRIMMDebug, and Bobby Whyte and Neil Brown for their feedback on early drafts of this paper. The first author also gives all praise and thanks to the Lord Jesus Christ for blessing him with the wonderful opportunity to conduct this research.
\end{acks}

\bibliographystyle{ACM-Reference-Format}
\bibliography{bibliography.bib, bibliography-local.bib, bibliography-with-last-accessed}

\end{document}